\documentclass[aps,prl,twocolumn,superscriptaddress]{revtex4-2}
\usepackage{graphicx}
\usepackage{dcolumn}
\usepackage{bm}
\usepackage[T1]{fontenc}
\usepackage[utf8]{inputenc}
\usepackage{color}
\usepackage{array}
\usepackage{varwidth}
\usepackage{amsmath}
\usepackage{booktabs}
\usepackage{amssymb}
\usepackage{graphicx}
\usepackage{multirow}
\usepackage[english,main=english]{babel}
\usepackage{hyperref}
\usepackage{float}
\usepackage{makecell}
\usepackage{array}
\definecolor{darkblue}{RGB}{0,0,139}

\newcolumntype{M}[1]{>{\centering\arraybackslash}m{#1}}
\hypersetup{
    colorlinks=True,
    linkcolor=blue,
    urlcolor=blue,
    citecolor=blue
}
\begin{document}

\title{Unconventional Superconductivity in $\mathrm{La_{3}Ni_{2}O_{7}}$ from the Perspective of Symmetry}

\author{Guan-Hao Feng}\email{fenggh@lingnan.edu.cn}
\affiliation{School of Physical Science and Technology, Lingnan Normal University, Zhanjiang, 524048, China}

\author{Jun Quan}\email{quanj@lingnan.edu.cn}
\affiliation{School of Physics, Changchun Normal University,  Changchun, 130032, China}
\affiliation{School of Physical Science and Technology, Lingnan Normal University, Zhanjiang, 524048, China}

\author{Yusheng Hou}
\affiliation{Guangdong Provincial Key Laboratory of Magnetoelectric Physics and Devices, Center for Neutron Science and Technology, School of Physics, Sun Yat-Sen University, Guangzhou, 510275, China}
\begin{abstract}
The recently discovered superconductor $\mathrm{La_{3}Ni_{2}O_{7}}$ has attracted significant attention due to its remarkably high transition temperature ($T_{c}$) under high pressure. Shortly after this discovery, thin-film $\mathrm{La_{3}Ni_{2}O_{7}}$ was demonstrated to exhibit ambient-pressure superconductivity; however, the corresponding $T_c$ is only about half that of the pressurized bulk material. This striking difference raises questions about the underlying mechanisms governing superconductivity in these two structures. To address this issue, we develop a phenomenological symmetry-based method to investigate the superconducting gap structure in $\mathrm{La_{3}Ni_{2}O_{7}}$. Using density-functional theory methods (DFT+$U$), together with the experimentally determined $T_c$ and structural symmetry, we find that both pressurized bulk and thin-film $\mathrm{La_{3}Ni_{2}O_{7}}$ exhibit $s_{\pm}$-wave pairing symmetry and two-gap superconductivity, yet their dominant microscopic pairing configurations are distinct. In the pressurized bulk, superconductivity is dominated by the out-of-plane pairing of the Ni-$d_{z^2}$ orbitals, while in the thin film, the in-plane pairing of the Ni-$d_{x^2-y^2}$ orbitals prevails. Furthermore, the observed reduction in $T_c$ can be attributed to this transition of the dominant pairing type, driven by the decreased ratio of inter-layer to intra-layer hoppings in the thin film. Our result sheds lights on the microscopic pairing in $\mathrm{La_{3}Ni_{2}O_{7}}$ and reveals the significance of the symmetry. This method can potentially be generalized to a broader range of unconventional superconductors.
\end{abstract}

\maketitle

\section{Introduction}
The bilayer Ruddlesden-Popper nickelate $\mathrm{La_{3}Ni_{2}O_{7}}$ exhibits a high superconducting onset temperature, which has attracted significant attention~\citep{sun_signatures_2023,wang_normal_2024}. However, experimental investigation of the superconducting pairing mechanism remains a major challenge due to the necessity of high-pressure condition~\cite{PhysRevB.111.104511}. More recently, thin-film $\mathrm{La_{3}Ni_{2}O_{7}}$ grown on $\mathrm{SrLaAlO_{4}}$ substrates has exhibited ambient-pressure superconductivity, providing a new platform for detailed experimental study~\cite{ko_signatures_2025,zhou_ambient-pressure_2025,liu_superconductivity_2025}. Notably, $T_c$ in thin films is about half that of the pressurized bulk~\cite{Yue_2025, ushio2025theoreticalstudyambientpressure}, and the reason for this reduction remains unclear. This naturally raises the question of the similarities and differences in the superconducting mechanisms between the two structures.

So far, considerable effort has been devoted to understanding the origin of the unconventional superconductivity and indicate that both pressurized bulk and thin-film $\mathrm{La_{3}Ni_{2}O_{7}}$ favor $s_{\pm}$-wave or $d$-wave pairing, driven by strong out-of-plane coupling or orbital hybridization~\cite{luo_bilayer_2023, yang_possible_2023, PhysRevLett.131.236002, PhysRevB.108.L201108, PhysRevB.108.L140504, PhysRevB.108.L180510,PhysRevLett.131.206501, PhysRevLett.133.146002,liao2024orbitalselectiveelectroncorrelationshightrm, PhysRevB.110.L060510, PhysRevB.108.165141,PhysRevB.109.165154, PhysRevB.110.104517, fan_superconductivity_2024, PhysRevLett.133.096002,PhysRevB.110.205122,Chen_electronic_2024,PhysRevLett.132.146002,sakakibara_possible_2024,jiang_high-temperature_2024,PhysRevLett.133.126501,PhysRevB.111.035108,yang_orbital-dependent_2024, PhysRevB.108.165141, PhysRevB.108.214522, Shen_2023, PhysRevLett.132.146002, PhysRevLett.132.036502, PhysRevB.111.174506}. Nevertheless, the role of symmetry in the pairing mechanisms remains subtle and complex. It is therefore intriguing to explore how symmetry influences superconductivity—or, perhaps more interestingly, to ask whether superconductivity can be induced by symmetry itself. To address this question, it is instructive to examine the most fundamental property shared by the two structures---symmetry. Comprehensive high-pressure experiments up to $104~\mathrm{GPa}$ have revealed that the superconducting transition in pressurized bulk $\mathrm{La_{3}Ni_{2}O_{7}}$ is accompanied by a structural transition from the $Amam$ to the $I4/mmm$ space group above $14~\mathrm{GPa}$~\cite{nwaf220,wang_structure_2024,wang_bulk_2024}. Meanwhile, thin-film $\mathrm{La_{3}Ni_{2}O_{7}}$ exhibits the $P4/mmm$ space group symmetry due to substrate-induced strain~\cite{ko_signatures_2025,zhou_ambient-pressure_2025,liu_superconductivity_2025}. Remarkably, the bilayer phases in both structures share the same layer group (LG) symmetry, $p4/mmm$, providing a unified symmetry framework for investigating their superconductivity.

Recent experiments have uncovered several key features of superconductivity in nickelates. First, the high $T_c$ remains robust under applied magnetic fields and shows little sensitivity to the low superconducting volume fraction~\citep{sun_signatures_2023,PhysRevB.109.144511,zhang_effects_2024,PhysRevX.14.011040,nwaf220,zhou_investigations_2025,hou_emergence_2023,zhang_high-temperature_2024}, suggesting a short coherence length. Second, the maximum superconducting volume fraction is approximately $41\%$, indicating that superconductivity predominantly arises from the bilayer phase~\citep{nwaf220}. Third, samples of varying quality consistently exhibit a linear temperature dependence of resistivity, indicative of strong electronic correlations and non-Fermi-liquid behavior~\citep{sun_signatures_2023,yuan_scaling_2022,zhang_high-temperature_2024}. These characteristics closely resemble those seen in cuprate and Fe pnictide superconductors. In cuprate superconductors, the superconducting gap primarily exhibits $d_{x^2-y^2}$-wave symmetry, typically described by the pairing harmonic $\cos k_x - \cos k_y$~\cite{can_high-temperature_2021}. By contrast, Fe-pnictide superconductors are widely believed to exhibit an $s_{\pm}$-wave gap, often parameterized by $s_{x^2y^2}$-wave harmonics $\cos k_x\cos k_y$, which change sign between electron and hole Fermi pockets~\cite{Hirschfeld_2011}. Moreover, it is well established that mean-field theories of strong coupling model can capture the key feature of the gap structure in cuprate and Fe-pnictide superconductors~\cite{Hirschfeld_2011, PhysRevLett.101.206404, can_high-temperature_2021}. For example, a Hubbard model with on-site repulsion and effective nearest-neighbor attraction can stabilize a $d_{x^2-y^2}$-wave superconducting state in a single CuO$_2$ monolayer~\cite{can_high-temperature_2021}. In this work, we generalize this approach to a multi-orbital case to study the superconductivity in La$_3$Ni$_2$O$_7$.

The effective interactions are strongly constrained and modulated by the underlying crystal symmetry and electronic structure, naturally favoring certain pairing harmonics. In this work, we develop a phenomenological, symmetry-based framework to analyze superconducting pairing in $\mathrm{La_{3}Ni_{2}O_{7}}$, without imposing a priori assumptions about the microscopic pairing configuration. Throughout this work, we focus on the possible superconducting states that preserve time-reversal symmetry (TRS) which can be identified by the eight one-dimensional (1D) irreducible representations of the point group $4/mmm$ ($D_{4h}$). Thus, TRS-breaking superconducting states, such as those with $s+id$-wave symmetry, are beyond the scope of this work. This method is applied to both pressurized bulk and thin-film $\mathrm{La_{3}Ni_{2}O_{7}}$. Our results reveal that, although both systems exhibit $s_{\pm}$-wave pairing symmetry, superconductivity in the pressurized bulk is dominated by the out-of-plane pairing of the $d_{z^2}$ orbitals, whereas in the thin film, it is governed by the in-plane pairing of the $d_{x^2-y^2}$ orbitals. This distinction explains why a reduced in-plane lattice constant in the thin film enhance $T_c$, whereas a reduced out-of-plane lattice constant does not~\cite{ko_signatures_2025}. Furthermore, our results suggest that the approximately halved $T_c$ in the thin film can be attributed to the transition of the dominant pairing type. Finally, our calculated spectra show closed agreement with angle-resolved photoemission spectroscopy (ARPES)~\cite{shen2025nodelesssuperconductinggapelectronboson} and scanning tunneling microscopy and spectroscopy (STM/STS) measurements~\cite{fan2025superconductinggapsrevealedstm}, indicating that symmetry may play an important role in superconductivity and that our symmetry-based approach may potentially be generalized to a wider range of superconductors.
 
 \section{Symmetry and superconductivity}

\begin{figure}[h]
 \centering{}\includegraphics[width=0.5\columnwidth,totalheight=1.0\columnwidth,keepaspectratio]{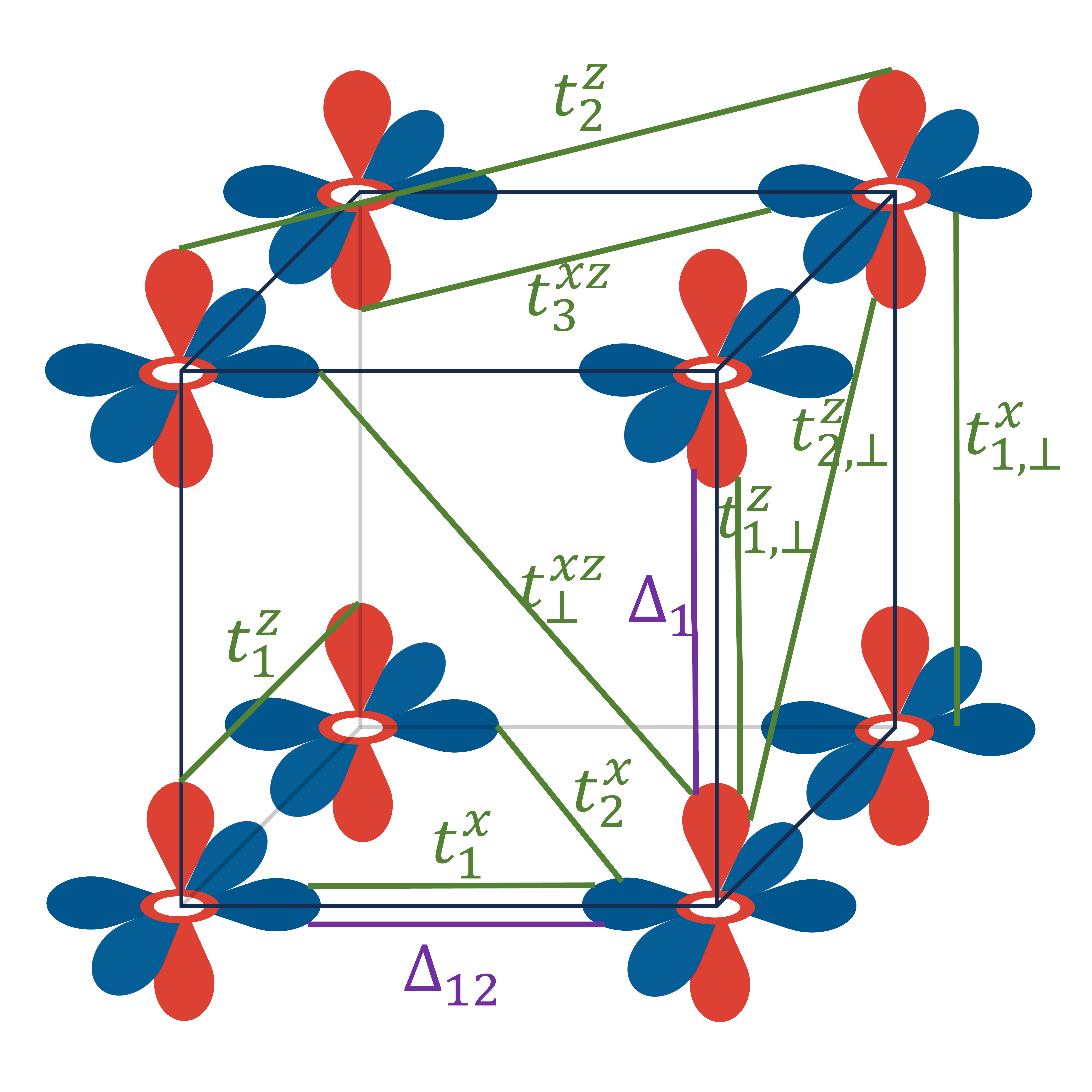}
\caption{
	Schematic of the hopping terms in the tight-binding model (Eq.~\ref{eq:2}). The two purple lines highlight the pairing types $\Delta_1$ and $\Delta_{12}$, identified as $\Delta_{\perp}^{z}$ ($s_{z^2}$-wave pairing) and $\Delta_{\parallel}^{x}$ ($s_{x^2+y^2}$-wave pairing), respectively, which are ultimately stabilized in our calculations.
	\label{fig: Figure1}}
\end{figure}

We start from a Hamiltonian on the lattice with strong on-site repulsion and short-range attraction~\cite{can_high-temperature_2021}, $H = H_{\text{kin}} + H_{\text{rep}} + H_{\text{attr}}$. Here $H_{\text{kin}}$, $H_{\text{rep}}$, and $H_{\text{attr}}$ denote the kinetic terms, on-site repulsion, and  short-range attraction, respectively. At the mean-field level, following the Bogoliubov–de Gennes (BdG) theory, this Hamiltonian can be cast into the familiar BdG form in the Nambu basis,
\begin{equation}
\mathcal{H}_{\boldsymbol{k}}^{\mathrm{BdG}} = \mathcal{H}_0 + \mathcal{H}_\Delta,\label{eq:1}
\end{equation}
with
$\mathcal{H}_0 = \mathrm{diag}\{\mathcal{H}_{\boldsymbol{k}}-\mu,\; -\mathcal{H}_{-\boldsymbol{k}}^{*}+\mu\},$
and
$\mathcal{H}_\Delta = \mathrm{offdiag}\{\tilde{\Delta}_{\boldsymbol{k}},\; \tilde{\Delta}_{\boldsymbol{k}}^\dagger\},$
where $\mathcal{H}_0$ corresponds to the normal part arising from $(H_{\text{kin}} + H_{\text{rep}})$ and $\mathcal{H}_\Delta$ encodes the pairing part generated by $H_{\text{attr}}$. We emphasize that this normal part is used as an effective non-superconducting-state reference entering the BdG Hamiltonian, rather than the true physical normal state realized above $T_c$. As we focus on TRS-preserving superconducting states, the normal part $\mathcal{H}_{\boldsymbol{k}} - \mu$ is required to be TRS-preserving as well. Concretely, the normal part is approximated by DFT+$U$ calculations followed by Wannier downfolding, yielding a two-orbital (Ni $d_{z^2}$ and $d_{x^{2}-y^{2}}$) convention-II tight-binding model on a bilayer square lattice, which respects the symmetries of the type-II magnetic layer group $p4/mmm1'$,

  \begin{align}
&\mathcal{H}_{\boldsymbol{k}}-\mu  \nonumber \\
&=T_{k}^{z}\rho_{0}\frac{\sigma_{0}+\sigma_{z}}{2}s_{0} \nonumber  +T_{k}^{x}\rho_{0}\frac{\sigma_{0}-\sigma_{z}}{2}s_{0}+V_{k}\rho_{0}\sigma_{x}s_{0} \nonumber \\
&+T_{k}'^{z}\rho_{x}\frac{\sigma_{0}+\sigma_{z}}{2}s_{0}+T_{k}'^{x}\rho_{x}\frac{\sigma_{0}-\sigma_{z}}{2}s_{0}+V_{k}'\rho_{x}\sigma_{x}s_{0}, \label{eq:2}
 \end{align}
 with
 \begin{align}
T_{k}^{x/z} &=\epsilon^{z}+2t_{1,\parallel}^{x/z}\left(\cos k_{x}+\cos k_{y}\right) + 4t_{2,\parallel}^{x/z}\cos k_{x}\cos k_{y}, \nonumber \\
 V_{k}&=2t_{1,\parallel}^{xz}(\cos k_{x}-\cos k_{y})\nonumber \\
T_{k}'^{x/z} &=t_{1,\perp}^{x/z} +2t_{2,\perp}^{x/z}, \left(\cos k_{x}+\cos k_{y}\right), \nonumber \\
V_{k}' &=2t_{1,\perp}^{xz}\left(\cos k_{x}-\cos k_{y}\right).  \nonumber
 \end{align}
Here, $\rho_{i}$, $\sigma_{i}$, and $s_{i}$ ($i=0,x,y,z$) represent the Pauli matrices acting on the layer, orbital, and spin degrees of freedom, respectively. The subscripts $\parallel$ and $\perp$ refer to intra-layer and inter-layer hoppings, respectively, while the superscripts $x$, $z$, and $xz$ denote hoppings originating from the $d_{x^{2}-y^{2}}$ orbital, the $d_{z^2}$ orbital, and orbital hybridization, respectively. The basis operators are defined as $\psi_{k\sigma}^{\dagger} = \left[c_{Az\uparrow}^{\dagger},\, c_{Az\downarrow}^{\dagger},\, c_{Ax\uparrow}^{\dagger},\, c_{Ax\downarrow}^{\dagger},\, c_{Bz\uparrow}^{\dagger},\, c_{Bz\downarrow}^{\dagger},\, c_{Bx\uparrow}^{\dagger},\, c_{Bx\downarrow}^{\dagger}\right]$, where $A$ and $B$ denote the two layers. We adopt the conventional lattice basis and include hopping terms up to next-nearest neighbors, as illustrated in Fig.~\ref{fig: Figure1}. This model is applicable to the bilayer phases in both the pressurized bulk and the thin-film structures. The distinction between these two cases arises solely from the choice of system-specific parameters~\citep{supplemental_material}. 

\begin{figure}[h]
 \centering{}\includegraphics[width=1\columnwidth,totalheight=1.0\columnwidth,keepaspectratio]{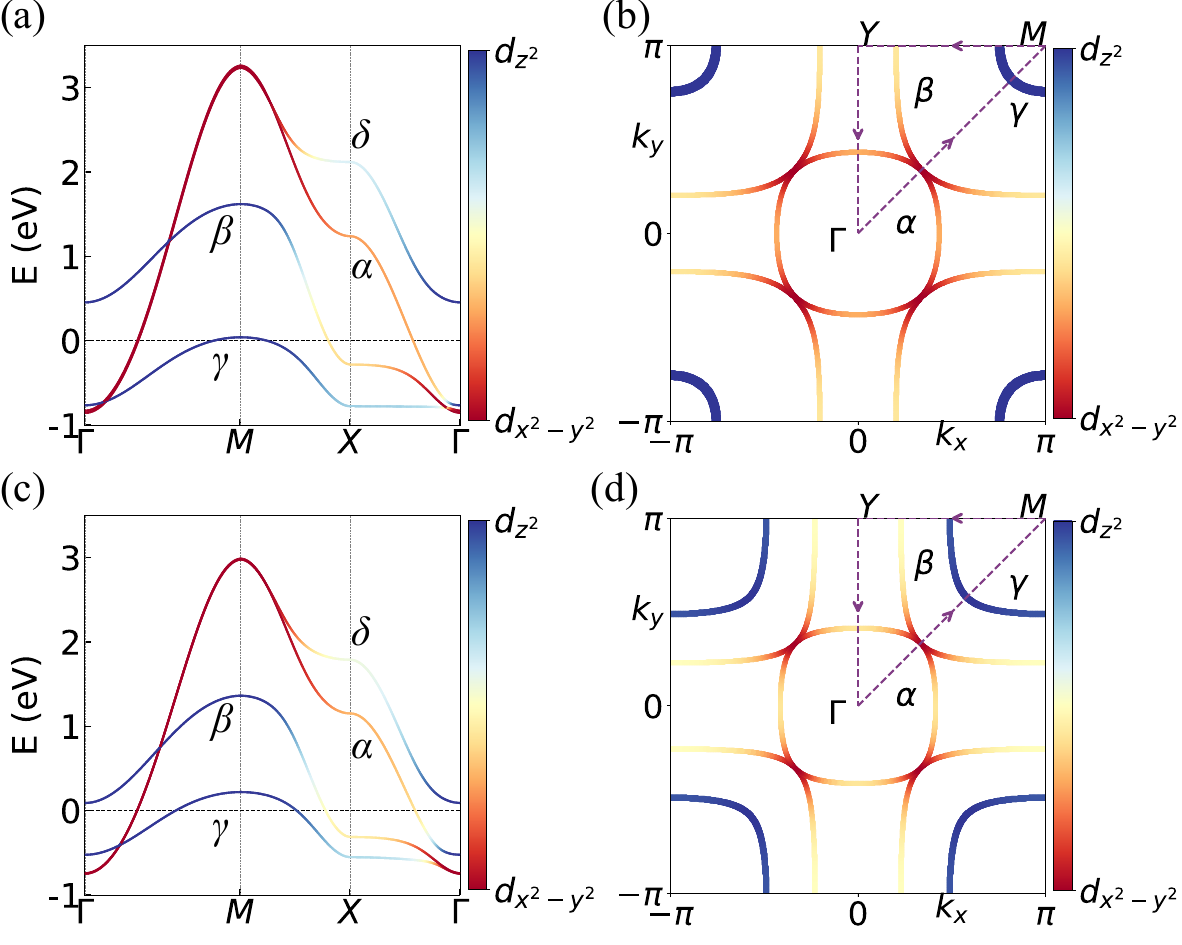}
\caption{
Tight-binding band structures and Fermi surfaces for (a), (b) the pressurized bulk and (c), (d) the thin film. The color bar here indicates the  weight of $d_{z^2}$ and $d_{x^2-y^2}$ orbitals. The specific parameter values are provided in Supplementary Notes~\cite{supplemental_material}.
	\label{fig: Figure2}}
\end{figure}

The band structures and Fermi surfaces are shown in Fig.~\ref{fig: Figure2}, where a larger $\gamma$ pocket can be observed in the thin film compared to the pressurized bulk. Orbital components are illustrated with color scales, indicating that the $\gamma$ sheet is mainly contributed by the $d_{z^2}$ orbital, while the $\alpha$ and $\beta$ sheets display a mixture of $d_{x^2-y^2}$ and $d_{z^2}$ orbitals in both cases. In cuprates, intra-orbital scattering of the $d_{x^2-y^2}$ orbital corresponds to intra-band scattering in single-band systems, resulting in $B_{1g}$ $d_{x^2-y^2}$-wave pairing. Notably, the shape of the $\beta$ sheet in the thin film closely resembles that of cuprates and may suggest the presence of $d_{x^2-y^2}$-wave pairing. However, the mixture of orbital components in the $\alpha$ and $\beta$ sheets (see Fig.~\ref{fig: Figure2}) indicates that, in multiband systems, intra-orbital scattering may be transformed into inter-band scattering, leading to gap sign changes between different Fermi surface sheets, i.e., $s_{\pm}$-wave pairing.
 
The short-range attractive interaction can be expressed as an extended multi-orbital attractive Hubbard interaction in a factorizable form, 
 \begin{align}
 & H_{\text{attr}}=
  \sum_{\boldsymbol{kk}'\alpha\beta ss'} V_{\text{eff}} c_{\boldsymbol{k}\alpha s}^{\dagger}c_{-\boldsymbol{k}\beta s'}^{\dagger}c_{-\boldsymbol{k}'\beta s'}c_{\boldsymbol{k}'\alpha s},\label{eq:3}
 \end{align}
 where $V_{\text{eff}}=\frac{-V_{\alpha\beta ss'}}{N}e^{-i(\boldsymbol{k}-\boldsymbol{k}')\cdot\boldsymbol{r}_{\alpha\beta ss'}}$, with  $s$ and $s'$ representing  the spin degrees of freedom, while $\alpha$ and $\beta$ denoting the degrees of freedom other than spin. Here, $N$ is the number of discrete momentum $\boldsymbol{k}$ (or $\boldsymbol{k}'$); $V_{\alpha\beta ss'}$ is the strength of the attractive potential; and the phase factor $e^{-i(\boldsymbol{k}-\boldsymbol{k}')\cdot\boldsymbol{r}_{\alpha\beta ss'}}$ arises from the Fourier transform of the Cooper pair creation operator $c_{\boldsymbol{k}\alpha s}^{\dagger}c_{-\boldsymbol{k}\beta s'}^{\dagger}$ and annihilation operator $c_{-\boldsymbol{k}'\beta s'}c_{\boldsymbol{k}'\alpha s}$. We then carry out the Hubbard–Stratonovich decoupling~\citep{supplemental_material} and obtain the mean-field pairing potential 
 \begin{align}
 H^{\text{MF}}_{\text{attr}} & =H_{\Delta}+\sum_{\alpha\beta ss'}\frac{N}{V_{\alpha\beta ss'}} |\Delta_{\alpha\beta ss'}|^{2},\label{eq:4}\\
 H_{\Delta} & =\sum_{\boldsymbol{k}\alpha\beta ss'}\left(\Delta_{\alpha\beta ss'}^{*}e^{i\boldsymbol{k}\cdot\boldsymbol{r}_{\alpha\beta}}c_{-\boldsymbol{k}\beta s'}c_{\boldsymbol{k}\alpha s}\right.\nonumber \\
 & \left.+\Delta_{\alpha\beta ss'}e^{-i\boldsymbol{k}\cdot\boldsymbol{r}_{\alpha\beta}}c_{\boldsymbol{k}\alpha s}^{\dagger}c_{-\boldsymbol{k}\beta s'}^{\dagger}\right).\label{eq:5}
 \end{align}
The combination of terms $\Delta_{\alpha\beta ss'}^{*}e^{i\boldsymbol{k}\cdot\boldsymbol{r}_{\alpha\beta ss'}}$ in $H_{\Delta}$ can give rise to various pairing symmetries, such as $s_{\pm}$, $p_{x}\pm ip_{y}$, $d_{x^{2}-y^{2}}$, $d_{xy}$, et al. The elements $\Delta_{\alpha\beta ss'}$ that share the same value collectively form the pairing matrix $\tilde{\Delta}_{i}=\Delta_{i}\gamma(k)\rho_{l}\sigma_{m}s_{n}$ in the Nambu basis. Thus, each $\tilde{\Delta}_i$ can be regarded as an independent type of pairing, whose pairing symmetry is determined by the pairing harmonic $\gamma(k)$. The corresponding free energy is then given by~\citep{supplemental_material, Coleman_2015,can_high-temperature_2021}
 \begin{equation}
 \mathcal{F}_{\mathrm{S}}(\Delta_i)=\frac{Nn}{V_{i}}\left|\Delta_i\right|^{2}-2k_{B}T\sum_{k\alpha}\ln\left[2\cosh(\beta E_{\boldsymbol{k}\alpha}^{\text{BdG}}/2)\right],\label{eq:6}
 \end{equation}
 where the eigenvalues of the BdG Hamiltonian are denoted as $E_{\boldsymbol{k}\alpha}^{\text{BdG}}$, and $\beta=1/k_{B}T$ is the inverse temperature. Here, $n$ is the number of nonzero elements in $\tilde{\Delta}_{i}$. When $\Delta_{i}=0$, Eq.~\ref{eq:6}   reduces to the non-superconducting free energy $\mathcal{F}_{\mathrm{N}}$.  The equilibrium gap value $\Delta_{i}$ can be identified by minimizing $\mathcal{F}_{\mathrm{S}}(\Delta_{i})$, which leads to the self-consistent BCS gap equation
 \begin{equation}
 \Delta_i=-\frac{V_{i}}{Nn}\sum_{\boldsymbol{k}}\text{Tr}\left[\frac{\partial\mathcal{H}_{\boldsymbol{k}}^{\text{BdG}}}{\partial\Delta_{i}}U_{\boldsymbol{k}}n_{F}(E_{\boldsymbol{k}}^{\text{BdG}})U_{\boldsymbol{k}}^{\dagger}\right],\label{eq:7}
 \end{equation}
 where $n_F$ is a diagonal matrix whose entries are the Fermi distribution functions for the eigenvalues $E_{\boldsymbol{k}}^{\text{BdG}}$. Here $U_{\boldsymbol{k}}$ is the unitary matrix that diagonalizes the BdG Hamiltonian. 
The favored pairing symmetry is determined by comparing the BCS condensation energy of each pairing type, $\mathcal{F}_{\mathrm{BdG},i} = \min\big(\mathcal{F}_{\mathrm{S}}(\Delta_{i})\big) - \mathcal{F}_{\mathrm{N}}$. In multiorbital superconductors, multiple coexisting pairing types may further lower the condensation energy. In this case, Eq.~\ref{eq:6} can generalize to the multi-pairing form by replacing the first term with $\sum_{i}\frac{Nn}{V_{i}}|\Delta_i|^{2}$. The optimal pairing configuration is found by minimizing $ \mathcal{F}_{\mathrm{S}}(\Delta_1, ..., \Delta_{n_{\mathrm{tot}}})$, which can be efficiently computed numerically. 
 
When TRS is preserved, each pairing type $\tilde{\Delta}_i$ should belong to one of the 1D pairing channels: $A_{1g}$, $A_{2g}$, $B_{1g}$, $B_{2g}$, $A_{1u}$, $A_{2u}$, $B_{1u}$, or $B_{2u}$ of $4/mmm$ ($D_{4h}$) point group~\citep{ono_symmetry_2019,PhysRevResearch.2.013064,ono_refined_2020}. Under any unitary operation $g \in 4/mmm$, the pairing matrix transforms as $U_{\boldsymbol{k}}(g)\tilde{\Delta}_{\boldsymbol{k},i}U_{-\boldsymbol{k}}^{T}(g) = \chi_{g}\,\tilde{\Delta}_{g\boldsymbol{k},i}$, with $\chi_g = \pm 1$, which uniquely identifies the 1D pairing channel of $\tilde{\Delta}_i$. Here $U_{\boldsymbol{k}}(g)$ is the matrix representation of unitary operation $g$ for the normal-part Hamiltonian. Given that generators for $p4/mmm$ are $\{4_{001}|0\}$, $\{2_{100}|0\}$, $\{2_{010}|0\}$, $\{2_{-110}|0\}$, and $\{-1|0\}$~\cite{supplemental_material}, the corresponding BdG unitary generators can be straightforwardly constructed by
\begin{equation}
	U_{\boldsymbol{k}}^{\mathrm{BdG}}(g)\equiv\left(\begin{array}{cc}
		U_{\boldsymbol{k}}(g)\\
		& \chi_{g}U_{-\boldsymbol{k}}^{*}(g)
	\end{array}\right).\label{eq:8}
\end{equation}
The anti-unitary symmetries, namely TRS $\mathcal{T}$ and particle--hole symmetry (PHS) $\mathcal{P}$, can also be constructed in a standard way~\cite{supplemental_material}. Here we focus on class DIII (characterized by $\mathcal{T}^2 = -1$ and $\mathcal{P}^2 = +1$) in the Altland--Zirnbauer classification, since class CII (with $\mathcal{T}^2 = -1$ and $\mathcal{P}^2 = -1$) is rarely realized in electronic systems~\citep{ono_refined_2020}. With the usage of these BdG generators, all of the symmetry-allowed pairing types can be identified systematically.

To determine which pairing types are favored, a straightforward approach is to compare the minimum free energy of each pairing type. However, before making such a comparison, we first need to determine the value of $V_i$ for each pairing type. This can be done by solving Eq.~\ref{eq:7} and choosing $V_i$ such that the experimental $T_c$ is reproduced. We note that not all values of $V_i$ are physically realizable in the system. Therefore, we compare the calculated $V_i$ with the corresponding hopping $|t|$ to screen out the infeasible pairing types. We then compare the minimum free energy of each feasible pairing type and identify the leading type. Finally, by selecting the pairing types that are feasible and compatible with the leading one, we obtain the ultimately stabilized pairing configuration.

In the procedure above, we have compared $V_i$ with the corresponding hopping $|t|$. This comparison is motivated by renormalization theory~\cite{yang_possible_2023}, where $V_i$ is related to the superexchange interaction via $V_i \sim \frac{3}{8} g_J J$, with $J \sim \frac{4 t^2}{\bar{U}}$, $g_J$ the renormalization factor, and $\bar{U} \sim U - J_H$. Although we do not explicitly compute $g_J$, comparing $V_i$ with $|t|$ still provides a useful criterion for assessing the potential emergence of pairing types. In this way, the structural symmetry, together with the experimental $T_c$, "selects" the ultimately stabilized pairing configuration.

\begin{table*}
	\begin{centering}
		\caption{List of symmetry-allowed mean-field superconducting pairing types in the $A_{1g}$
			pairing channel. Here the Pauli matrix $\tau_{i}$ denotes the particle-hole freedom.  \label{tab: A1g}}
		\par\end{centering}
	\centering{}%
	\begin{tabular}{cc}
		\hline 
		\multicolumn{2}{c}{$\mathcal{H}_{\Delta}$ in the $A_{1g}$ pairing channel}\tabularnewline
		\hline 
		$\Delta_{0}\tau_{y}\rho_{0}(\frac{\sigma_{0}+\sigma_{z}}{2})s_{y}$ & $\Delta_{8}\left[\left(\cos k_{x}+\cos k_{y}\right)\tau_{y}\rho_{0}(\sigma_{0}+\sigma_{z})s_{y}\right]$\tabularnewline
		$\Delta_{1}\tau_{y}\rho_{x}(\frac{\sigma_{0}+\sigma_{z}}{2})s_{y}$ & $2\Delta_{9}\left[\left(\cos k_{x}-\cos k_{y}\right)\tau_{y}\rho_{0}\sigma_{x}s_{y}\right]$\tabularnewline
		$\Delta_{2}\tau_{y}\rho_{0}(\frac{\sigma_{0}-\sigma_{z}}{2})s_{y}$ & $\Delta_{10}\left[\left(\cos k_{x}+\cos k_{y}\right)\tau_{y}\rho_{x}(\sigma_{0}+\sigma_{z})s_{y}\right]$\tabularnewline
		$\Delta_{3}\tau_{y}\rho_{x}(\frac{\sigma_{0}-\sigma_{z}}{2})s_{y}$ & $2\Delta_{11}\left[\left(\cos k_{x}-\cos k_{y}\right)\tau_{y}\rho_{x}\sigma_{x}s_{y}\right]$\tabularnewline
		$\Delta_{4}\left[\sin k_{x}\tau_{y}\rho_{z}(\sigma_{0}+\sigma_{z})s_{0}-\sin k_{y}\tau_{x}\rho_{z}(\sigma_{0}+\sigma_{z})s_{z}\right]$ & $\Delta_{12}\left[\left(\cos k_{x}+\cos k_{y}\right)\tau_{y}\rho_{0}(\sigma_{0}-\sigma_{z})s_{y}\right]$\tabularnewline
		$2\Delta_{5}\left(\sin k_{x}\tau_{y}\rho_{z}\sigma_{x}s_{0}+\sin k_{y}\tau_{x}\rho_{z}\sigma_{x}s_{z}\right)$ & $\Delta_{13}\left[\left(\cos k_{x}+\cos k_{y}\right)\tau_{y}\rho_{x}\left(\sigma_{0}-\sigma_{z}\right)s_{y}\right]$\tabularnewline
		$2\Delta_{6}\left(\sin k_{x}\tau_{y}\rho_{y}\sigma_{y}s_{0}+\sin k_{y}\tau_{x}\rho_{y}\sigma_{y}s_{z}\right)$ & $4\Delta_{14}\sin k_{x}\sin k_{y}\tau_{x}\rho_{0}\sigma_{y}s_{x}$\tabularnewline
		$\Delta_{7}\left[\sin k_{x}\tau_{y}\rho_{z}(\sigma_{0}-\sigma_{z})s_{0}-\sin k_{y}\tau_{x}\rho_{z}(\sigma_{0}-\sigma_{z})s_{z}\right]$ & $4\Delta_{15}\sin k_{x}\sin k_{y}\tau_{x}\rho_{x}\sigma_{y}s_{x}$\tabularnewline
		\hline 
	\end{tabular}
\end{table*}

\section{Numerical calculation}
\subsection{Pressurized  Bulk}
As a representative, let's implement the model and method on the pressurized bulk under $29.6$~GPa (with $a = b = 3.68$~\AA\ and $c = 19.38$~\AA), whose $T_{c}\approx 80~\mathrm{K}$~\cite{nwaf220}. The normal part of the BdG Hamiltonian can be obtained via DFT+$U$ calculations and Wannier downfolding (see the Supplemental Notes for details~\cite{supplemental_material}). In our analysis, we adopt $U = 3.5~\mathrm{eV}$ for the pressurized bulk, following Ref.~\cite{yang_orbital-dependent_2024}, where the resulting $\gamma$ band is in good agreement with the experimental observations.

First, we use the BdG generators from Eq.~\ref{eq:8} to systematically identify all symmetry-allowed pairing types by the Qsymm software package and consider the short-range pairing terms up to the next-nearest neighbors \cite{varjas_qsymm_2018}. The $A_{1g}$-channel pairing types $\mathcal{H}_{\Delta}$ are listed in Tab.~\ref{tab: A1g}; results for other channels are given in the Supplementary Notes~\cite{supplemental_material}. We first focus on the Pauli matrix structure of each pairing type, which provides insight into their microscopic origins. Terms with $\rho_{0,z(x,y)}$ denote intra-layer (inter-layer) pairing, while $\sigma_{0}+\sigma_{z}$ ($\sigma_{0}-\sigma_{z}$) identifies pairing within the $d_{z^2}$ ($d_{x^{2}-y^{2}}$) orbitals. Pairing with $\sigma_{x,y}$ indicate hybridization between the $d_{z^2}$ and $d_{x^{2}-y^{2}}$ orbitals. Notably, some pairing types share identical layer and orbital characters but exhibit different pairing symmetries. For instance, both $\Delta_{7}$ and $\Delta_{12}$ correspond to intra-layer pairing of $d_{x^2-y^2}$ electrons; however, $\Delta_{7}$ exhibits $p$-wave symmetry; whereas $\Delta_{12}$ possesses $s_{x^2+y^2}$-wave symmetry.

\begin{figure}
	\centering
	\begin{centering}
		\includegraphics[width=1\columnwidth,totalheight=1.5\columnwidth,keepaspectratio]{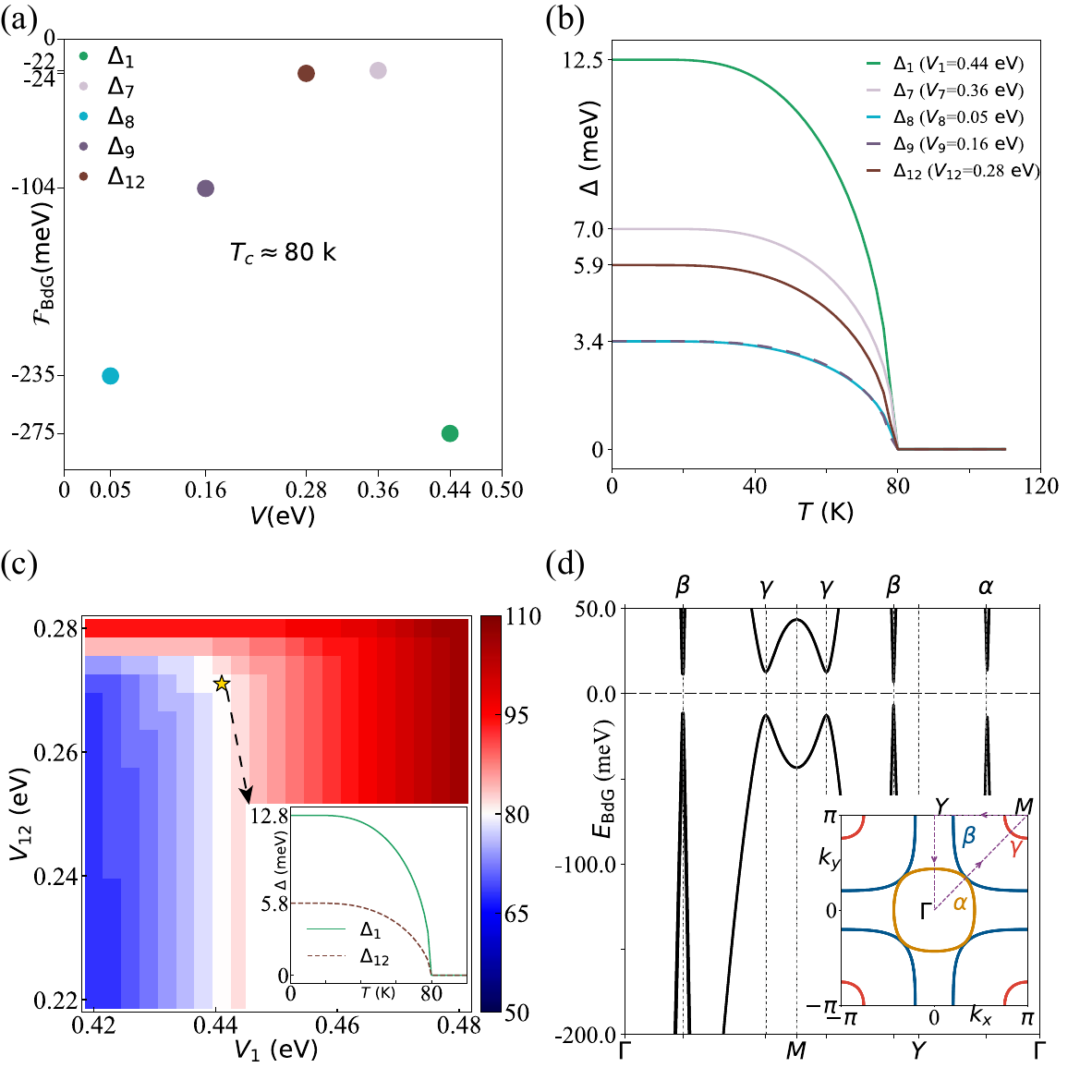}
		\par\end{centering}
	\centering{}
	\caption{
	(a) Comparison of the BCS condensation energy $\mathcal{F}_{\mathrm{BdG}}$ and the pairing strength $V_{i}$ for each feasible pairing type in the pressurized bulk under the constraint of $T_c = 80~\mathrm{K}$.
	(b) Temperature dependence of each $\Delta_{i}$ obtained by using the corresponding $V_i$ values in panel (a).
	(c) Superconducting transition temperature $T_c$ as a function of $V_{1}$ and $V_{12}$ when both interactions are present. The white region indicates where $T_c \approx 80~\mathrm{K}$, and the color scale shows the value of $T_c$. The inset displays the temperature dependence of $\Delta_{1}$ and $\Delta_{12}$ for the parameter set marked by the yellow star.
	(d) Schematic illustration of the momentum path and the corresponding energy spectrum, with $\{\Delta_{1}, \Delta_{12}\} = \{12.8, 5.8\}$~meV, as taken from the yellow star in (c), corresponding to $\Delta_{\perp}^{z}$ ($s_{z^2}$-wave pairing) and $\Delta_{\parallel}^{x}$ ($s_{x^2+y^2}$-wave pairing), respectively. Along the $(0,0)-(\pi,\pi)$ direction, a superconducting gap of approximately $11.3~\mathrm{meV}$ is observed at the $\beta$ pocket, and $12.8~\mathrm{meV}$ at the $\gamma$ pocket.
	\label{fig: Figure3}
	}
\end{figure}

Next, we choose the parameter value $V_i$ that reproduces the experimentally measured $T_c$ for each pairing type. Specifically, for the $A_{1g}$ channel, we find that only the five pairing types satisfying $V_{i}<|t|$, where $|t|$ is the corresponding hopping parameter. These pairing types are thus identified as feasible, i.e., $\{\Delta_{1}, \Delta_{7}, \Delta_{8}, \Delta_{9}, \Delta_{12}\}$, as shown in Fig.~\ref{fig: Figure3}(a). The minimization of free energy in Eq.~\ref{eq:6} for each feasible pairing type can then be carried out using the differential evolution method and cross-validated by self-consistently solving Eq.~\ref{eq:7}. Notably, $\Delta_1$ in the $A_{1g}$ channel yields the lowest BCS condensation energy among all of the pairing types in all channels, indicating that the leading pairing type in pressurized bulk belongs to the $A_{1g}$ channel with $s_{z^2}$-wave symmetry. We also verify $T_c$ of each feasible pairing type in Fig.~\ref{fig: Figure3}(b), which shows that the $V_i$ values obtained in Fig.~\ref{fig: Figure3}(a) indeed reproduce experimental transition temperature.

Finally, to examine coexistence or competition, we take $\Delta_{1}$ as the leading type and minimize the free energy with respect to pairs involving $\Delta_{1}$ and each feasible type. The results show that $\Delta_{8}$ and $\Delta_{9}$ compete with $\Delta_{1}$ and cannot coexist, while $\Delta_{7}$ and $\Delta_{12}$ are compatible and can be stabilized together with $\Delta_{1}$. This distinction arises from their electronic structures: $\Delta_{1}$ and $\Delta_{8}$ share the same Pauli matrix structure, causing competition, and $\Delta_{9}$ involves orbital hybridization, which prevents further free energy reduction when coexisting with $\Delta_{1}$. Furthermore, as shown in Fig.~\ref{fig: Figure3}(a), $\mathcal{F}_{\mathrm{BdG},12} < \mathcal{F}_{\mathrm{BdG},7}$ under the constraint of $T_c = 80 \text{K}$, indicating $\Delta_{12}$ is more favorable than $\Delta_{7}$. Therefore, $\Delta_{1}$ and $\Delta_{12}$ (i.e., $s_{z^2}$ pairing $\Delta_{\perp}^{z}$ and $s_{x^2+y^2}$-wave pairing $\Delta_{\parallel}^{x}$) are ultimately stabilized, supporting robust $s_{\pm}$-wave pairing symmetry in the pressurized bulk~\cite{sakakibara_possible_2024}.

In Fig.~\ref{fig: Figure3}(c), we present the dependence of the transition temperature $T_c$ on $\{V_{1}, V_{12}\}$ when both $\Delta_{1}$ and $\Delta_{12}$ are simultaneously present. As a representative case, for the parameter point marked by the star, the coherence between these two pairing channels results in an identical $T_c$. Moreover, the vertical stripe-like features in Fig.~\ref{fig: Figure3}(c) indicate that $\Delta_{1}$ is the dominant pairing component and plays a primary role in determining $T_c$. Using the pairing order parameters obtained from Fig.~\ref{fig: Figure3}(c), we find that the  energy spectrum is fully gapped along the high-symmetry path, as depicted in Fig.~\ref{fig: Figure3}(d). In addition, pronounced coherence peaks are clearly visible on the $\alpha$, $\beta$, and $\gamma$ Fermi surfaces.

\subsection{ Thin film }
\begin{figure}
	\centering
	\includegraphics[width=1\columnwidth,totalheight=1\columnwidth,keepaspectratio]{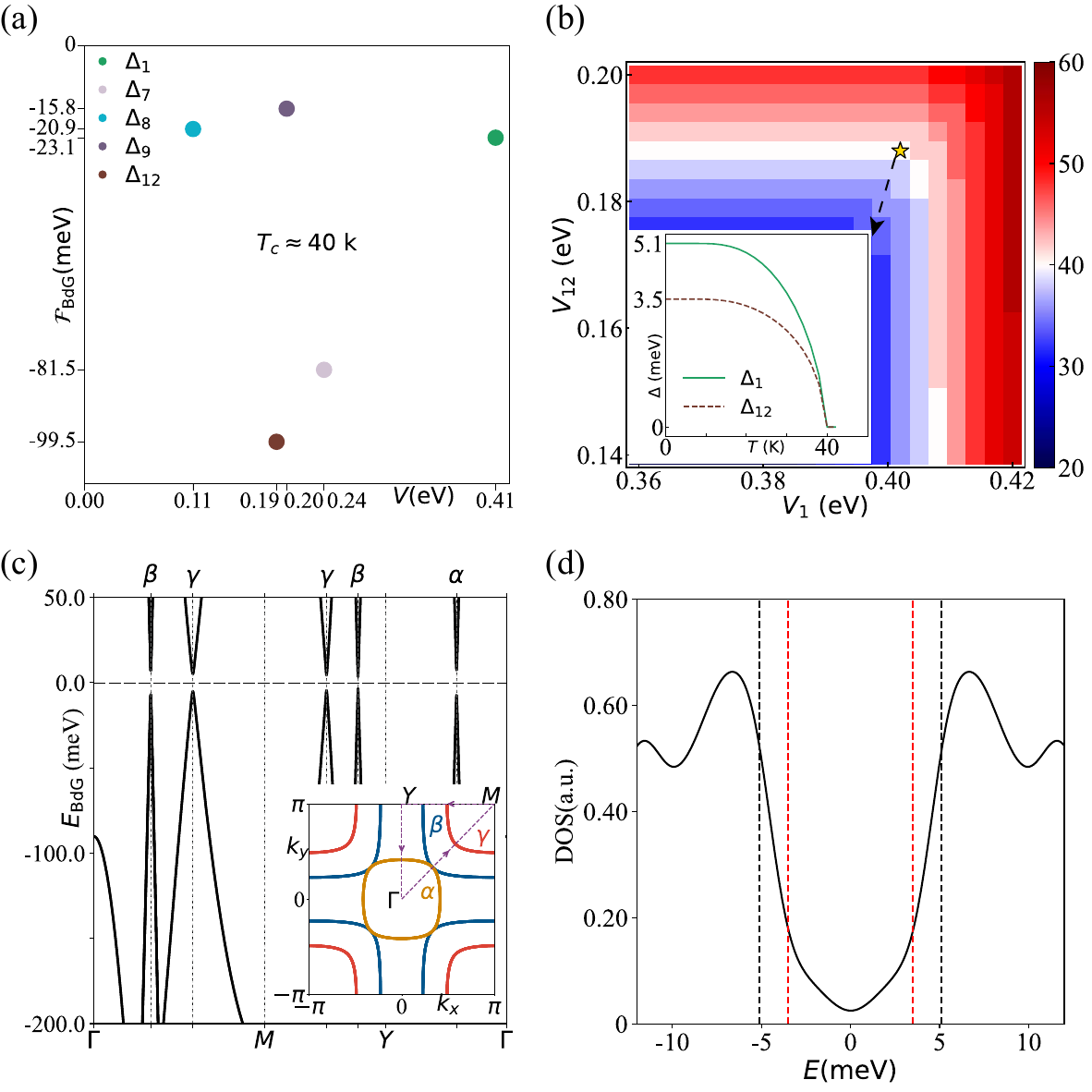}
	\caption{
	(a) Campraison of the BCS condensation energy $\mathcal{F}_{\mathrm{BdG}}$ and $V_{i}$ of each feasible pairing type for the thin film under the constraint of $T_c = 40 \text{K}$
	(b) Transition temperature $T_c$ as a function of $V_{1}$ and $V_{12}$. White regions indicate parameter regimes with $T_c \approx 40~\mathrm{K}$; the color scale depicts $T_c$. The inset shows the temperature dependence of $\Delta_{1}$ and $\Delta_{12}$ for the parameter set marked by the yellow star.
	(c) Energy spectrum along the momentum path shown in the inset, with $\Delta_{1}$ and $\Delta_{12}$ taken from the yellow star in (b), i.e., $\{\Delta_{1},\Delta_{12}\} = \{5.1, 3.5\}$~meV. The energy gap is approximately $7.2$~meV at the $\beta$ band and $5.3$~meV at the $\gamma$ band. This momentum path enables direct comparison with the experimental results reported in Ref.~\cite{shen2025nodelesssuperconductinggapelectronboson}.
	(d) Normalized density of states (DOS), for which the integral over the entire energy range is unity, enabling direct comparison with the experimental results reported in Ref.~\cite{fan2025superconductinggapsrevealedstm}. The black and red dashed lines denote $\{\Delta_{1},\Delta_{12}\}$  obtained in our calculations, respectively.
	\label{fig: Figure4}}
\end{figure}

For the normal part of the BdG Hamiltonian for the half-UC thin film, we adopt the  Sr-doping model  reported in Ref.~\cite{Yue_2025} ($a = b = 3.75$~\AA\ and $c = 20.82$~\AA), where we consider the hoppings up to next-nearest neighbors. We now proceed with the approach to investigate the superconducting properties. The matrix representations of the symmetry-allowed pairing types are identical to those in the pressurized bulk. Given that experimental measurements indicate $T_{c}$ varies from approximately $26$~K to $42$~K~\cite{ko_signatures_2025}, we adopt $T_c = 40$~K as a representative value in our calculations. By comparing the minimum free energy of each pairing type, we find that the leading pairing type also belong to the $A_{1g}$ channel with $s_{\pm}$-wave symmetry and the feasible pairing types are identical to the bulk case~\cite{supplemental_material}, as shown in Fig.~\ref{fig: Figure4}(a). Following the competitive relationships discussed in the pressurized bulk, the ultimately stabilized pairing types are again $\{\Delta_{1}, \Delta_{12}\}$.  

The key distinction, however, is that the in-plane pairing $\Delta_{12}$ (i.e., $s_{x^2+y^2}$-wave pairing $\Delta_{\parallel}^{x}$) now emerges as the leading instability in terms of free energy. Notably, as shown in Fig.~\ref{fig: Figure4}(b), when we focus on the parameter region under the constraint of $T_{c} = 40~\mathrm{K}$, the vertical segment of the $\{V_{1}, V_{12}\}$ regime is narrow at $V_{1} = 0.408~\mathrm{eV}$, which is close to $|t_{1,\perp}^{z}| = 0.439~\mathrm{eV}$. This suggests that such an interaction strength is unlikely to be realized in real materials. Therefore, we conclude that the in-plane pairing $\Delta_{12}$ plays a dominant role in determining $T_{c}$, as reflected by the horizontal stripe-like region. This dominant pairing-type transition can be attributed to the decreased ratio $|t_{1,\perp}^{z}/t_{1,\parallel}^{x}|$ in the thin film. This result is consistent with experimental observations that a reduced in-plane lattice constant leads to an enhanced $T_c$, whereas a reduced out-of-plane lattice constant does not~\cite{ko_signatures_2025}. Due to the weaker in-plane attractive interaction strength in the $d_{x^2-y^2}$ orbitals compared to the stronger interlayer attractive interaction in the $d_{z^2}$ orbitals, $T_c$ in the thin film is lower than that in the pressurized bulk. Therefore, the reduction in $T_c$ observed in the thin film can be primarily attributed to the transition in the dominant pairing type. 

In Fig.~\ref{fig: Figure4}(c), we present the calculated spectrum, which reveals a pronounced waterfall-like feature near the Fermi surface. The energy spectrum is fully gapped along the high-symmetry path, in close agreement with ARPES measurements on superconducting (La,Pr,Sm)$_3$Ni$_2$O$_7$ thin films (2 or 3 UC) reported in Ref.~\cite{shen2025nodelesssuperconductinggapelectronboson}. We further calculate the normalized density of states (DOS) and find the "V-shape" nodal gap signature near zero bias, with a clear two-gap feature observed, as shown in Fig.~\ref{fig: Figure4}(d). This result is in good correspondence with the tunneling spectra obtained from STM/STS experiments~\cite{fan2025superconductinggapsrevealedstm}. Although ARPES measurements report a gap size of approximately $18~\mathrm{meV}$ and STM/STS coherence peaks are located near $\pm20~\mathrm{meV}$—both somewhat larger than our calculated values—this discrepancy can likely be attributed to the omission of higher-order pairing interactions in our theoretical treatment.

For completeness, we now discuss the behavior of $d$-wave pairing in the thin film, focusing on two 1D irreducible representations and their commonly discussed pairing symmetries: $B_{1g}$ ($d_{x^2-y^2}$) and $B_{2g}$ ($d_{xy}$). These pairing types have been systematically identified in our framework using the BdG generators, namely $B_{1g}$ ($\Delta_{10}$) and $B_{2g}$ ($\Delta_{10}$)~\cite{supplemental_material}.
In the thin-film case, our calculations show that the minimum free energy for $B_{1g}$ ($\Delta_{10}$) pairing is approximately $-0.008$~eV, which is substantially lower than that for $A_{1g}$ ($\Delta_{12}$). Similarly, for $B_{2g}$ ($\Delta_{10}$) pairing, the relevant hopping parameter $|t_{2,\parallel}^x|$ is relatively small, and its minimum free energy (around $-0.057$~eV) is also considerably lower than that of $A_{1g}$ ($\Delta_{12}$)~\cite{supplemental_material}. Thus, both types of $d$-wave pairing are excluded as leading pairing instabilities in this system.
Examining the projected gap function on the Fermi surfaces further reveals that these differences are closely related to the presence of stronger inter-band scattering processes compared to intra-band scattering within the $\beta$ sheet. Such inter-band effects suppress the $d$-wave pairing channel and favor $s_{\pm}$-wave symmetry instead. We also note that the minimum free energy of $\Delta_{12}$ in the $A_{1g}$ channel is slightly lower than those ($\approx -0.097$~eV) of spin-triplet $p+ip$-wave pairing types $\Delta_6$ in the $A_{1u}$, $A_{2u}$, $B_{1u}$, and $A_{2u}$ channels~\cite{supplemental_material}. This difference becomes more pronounced when the $k$-summation in Eq.~\ref{eq:7} is performed over a denser momentum mesh, further indicating that the instability to spin-singlet pairing $\Delta_{12}$ dominates over those to spin-triplet pairing.

\section{Comparison with RPA calculations and pairing mechanism discussion}
\begin{figure}
 \centering{}\includegraphics[width=1\columnwidth,totalheight=1.0\columnwidth,keepaspectratio]{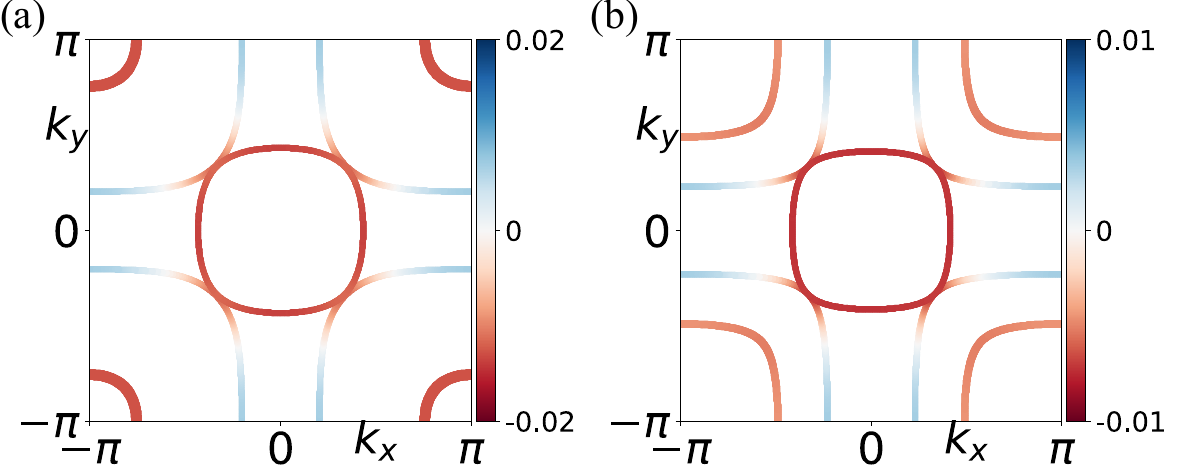}
\caption{
	Distribution of the projected gap function for $\tilde{\Delta}_{\text{tot}}=\tilde{\Delta}_{1}+\tilde{\Delta}_{12}$ on the Fermi surfaces for (a) the pressurized bulk and (b) thin film . Here the coefficients $\{\Delta_1,\Delta_{12}\}$ are set to the same values as in Figs.~\ref{fig: Figure3} and \ref{fig: Figure4}, respectively.}
	\label{fig: Figure5}
\end{figure}

It is necessary to compare our results with those obtained via random phase approximation (RPA) calculations, which are widely regarded as an effective method for studying unconventional superconductivity in correlated electronic systems. A straightforward way to make this comparison is to project the pairing matrices $\tilde{\Delta}_{i}$ onto the Fermi surface using the band eigenvectors $U_{a}$ of $\mathcal{H}_{\boldsymbol{k}} - \mu$ in Eq.~\ref{eq:1}. Here, $U_{a}$ is an $8\times2$ matrix owing to the inclusion of spin degrees of freedom, and $a = \alpha, \beta, \gamma, \delta$ labels the four bands. Given the form of the BdG Hamiltonian, the projected gap matrices on the Fermi surfaces are then given by
\begin{equation}
    \tilde{\Delta}_{aa}^{\mathrm{proj}}(\boldsymbol{k}_{F}) = U_{a}^\dagger(\boldsymbol{k}_{F})\, \tilde{\Delta}(\boldsymbol{k}_{F})\, U_{a}^*(-\boldsymbol{k}_{F})\,,
\end{equation}
where $\boldsymbol{k}_{F}$ denote the momentum points on the Fermi surface. For spin-singlet pairing types such as $\tilde{\Delta}_1$ and $\tilde{\Delta}_{12}$ in the $A_{1g}$ channel, each projected gap matrix $\tilde{\Delta}^{\mathrm{proj}}_{aa}$ is a $2\times2$ off-diagonal matrix with elements of opposite amplitude, reflecting the singlet spin configuration. We select the upper off-diagonal component as the gap function and illustrate the distribution in Fig.~\ref{fig: Figure5}. Both the pressurized bulk and thin film exhibit $s_{\pm}$-wave symmetry: the $\alpha$ and $\gamma$ pockets are fully gapped with the same sign, whereas the gaps on the $\beta$ pockets are predominantly of the opposite sign and vanish at accidental nodal points,  where the $\beta$ and $\alpha$ pockets approach each other.

For the pressurized bulk, as shown in Fig.~\ref{fig: Figure5}(a), the gap distribution is consistent with the RPA results reported in Ref.~\cite{zhang_structural_2024}, where the gap amplitude on the $\alpha$ sheet is comparable to that on the $\gamma$ sheet. For the thin film, since the tight-binding parameters in this work are taken from Ref.~\cite{Yue_2025}, the close agreement between our results in Fig.~\ref{fig: Figure5}(b) and the RPA calculations shown in Ref.~\cite{Yue_2025} further corroborates the validity of our method. Notably, we found that the gap amplitude on the $\alpha$ sheet is noticeably larger than that on the $\gamma$ sheet. The possible mechanism underlying the difference between the pressurized bulk and the thin film is discussed below.

The specific forms of the ultimately stabilized pairing, $\Delta_1$ and $\Delta_{12}$, could provide insight into the underlying mechanism. Within the framework of spin-fluctuation-mediated pairing, the term $\tilde{\Delta}_1 = \Delta_1 \tau_y \rho_x \left( \frac{\sigma_0 + \sigma_z}{2} \right) s_y$ originates from intra-orbital scattering within the $d_{z^2}$ orbital, whereas $\tilde{\Delta}_{12} = \Delta_{12}\left[(\cos k_x+\cos k_y) \tau_y \rho_0 (\sigma_0 - \sigma_z) s_y\right]$ arises from intra-orbital scattering in the $d_{x^2-y^2}$ orbital.
Although both terms result from intra-orbital scattering, the hybridization of orbital components in the $\alpha$ and $\beta$ sheets enables inter-band scattering. Notably, the degree of $d_{z^2}$ orbital mixing is stronger in the thin film compared to the pressurized bulk, as shown in Fig.~\ref{fig: Figure2}. Our results show that $V_{1} > V_{12}$ in both the pressurized bulk and thin film, indicating that intra-orbital scattering in the $d_{z^2}$ orbital is stronger than that in the $d_{x^2-y^2}$ orbital. Consequently, the enhanced $d_{z^2}$ orbital mixing in the thin film increases the inter-band scattering between the $\alpha$ and $\beta$ sheets, resulting in a relatively large amplitude of the superconducting gap on the $\alpha$ sheet.
Furthermore, the reduced ratio $|t_{1,\perp}^{z}/t_{1,\parallel}^{x}|$ in the thin film leads to a smaller $V_{1,\perp}^{z} / V_{1,\parallel}^x$ (i.e., $V_{1}/V_{12}$) compared to the pressurized bulk, indicating that intra-orbital scattering within the $d_{x^2-y^2}$ orbital is relatively enhanced. As a result, the in-plane $s_{x^2+y^2}$-wave pairing ($\Delta_{\parallel}^{x}$) is promoted as the leading instability as shown in Fig.~\ref{fig: Figure4}(a).
 It is worth emphasizing that different model may lead to different the pairing results, depending on the details of doping. For example, in Ref.~\cite{Yue_2025}, Sr is doped only into the middle La layer of the La$_3$Ni$_2$O$_7$ thin film, which favors $s_{\pm}$-wave pairing. In contrast, a hole-doped La$_3$Ni$_2$O$_7$ thin film may instead exhibit $d$-wave pairing, as reported in Ref.~\cite{zhang2025compressivestrainturnsspm}.

\section{Conclusion and Discussion}
In this work, we present a phenomenological symmetry-based approach to investigate the superconducting gap structure in La$_3$Ni$_2$O$_7$, including the dominant pairing symmetry, size and shape of gap, and microscopic orbital and layer configuration. We find that two types of pairing are likely to emerge simultaneously: out-of-plane $s_{z^2}$-wave pairing of the $d_{z^2}$ orbitals ($\Delta_{\perp}^{z}$) and in-plane $s_{x^2+y^2}$-wave pairing of the $d_{x^2-y^2}$ orbitals ($\Delta_{\parallel}^x$). 
For pressurized bulk, the out-of-plane pairing $\Delta_{\perp}^{z}$ dominates and primarily determines $T_c$. In contrast, in the thin-film case, the dominant pairing type shifts to the in-plane $s_{x^2+y^2}$-wave pairing $\Delta_{\parallel}^{x}$, consistent with experimental observations that a smaller in-plane lattice constant leads to a higher $T_c$, whereas a smaller out-of-plane lattice constant does not~\cite{ko_signatures_2025}. Our findings suggest that the approximately halved $T_c$ in thin films can be attributed to this transition in the dominant pairing type. 
We then calculate the energy spectra for both cases, finding that the gap sizes and their trends are in close agreement with ARPES and STM/STS measurements~\cite{shen2025nodelesssuperconductinggapelectronboson,fan2025superconductinggapsrevealedstm}. We note that, although our results indicate a fully gapped superconducting state along the high-symmetry path in both cases, accidental nodes can still occur due to the topological properties of the $A_{1g}$ channel~\cite{bzdusek_robust_2017}, depending on the specific model considered. Our results are robust against small perturbations in the hopping parameters.

Finally, we compare our results with those obtained from RPA calculations. For the pressurized bulk, the distribution of the gap function is consistent with that reported in Ref.~\cite{zhang_structural_2024}. For the thin film, in particular, by employing the same tight-binding model, we find that our results are in good agreement with those in Ref.~\cite{Yue_2025}, further demonstrating that inter-band scattering may give rise to $s_{\pm}$-wave pairing. The reduced ratio $|t_{1,\perp}^{z}/t_{1,\parallel}^{x}|$ in the thin film may enhance the intra-orbital scattering within the $d_{x^2-y^2}$ orbital, thereby promoting the in-plane $s_{x^2+y^2}$-wave pairing ($\Delta_{\parallel}^{x}$) as the leading instability.

This work can be regarded as a phenomenological yet symmetry-guided BCS analysis that relies on the experimentally observed $T_c$. Nevertheless, if one is primarily interested in possible TRS–preserving superconducting states in La$_3$Ni$_2$O$_7$, our method is able to capture the key features of the gap structure. Moreover, our unified framework can be straightforwardly applied to both high-pressure bulk and thin-film La$_3$Ni$_2$O$_7$ systems. This symmetry-based framework is applicable not only to bilayer La$_3$Ni$_2$O$_7$, but also to other unconventional superconductors with well-defined space groups. It highlights the close relationship between symmetry and superconductivity, and may help rationalize why the emergence of superconductivity is often associated with high-symmetry crystalline environments. A more detailed exploration of this connection is left for future work.

\begin{acknowledgments}
 We thank Zhongbo Yan, Cui-Qun Chen and Zhi-hui Luo for helpful
 discussions. G.F. thanks Seishiro Ono for valuable discussions. This work is supported by Guangdong Basic and Applied Basic Research Foundation (Grant No.~2023A1515110002, 2023A1515011796, 2024A1515011908) and the National Natural Sciences Foundation of China (Grant No.~12474247 and 92165204). Yusheng Hou acknowledges the support from Guangdong Provincial Key Laboratory of Magnetoelectric Physics and Devices (Grant No. 2022B1212010008) and Research Center for Magnetoelectric Physics of Guangdong Province (Grants 2024B0303390001).
\end{acknowledgments}

\bibliography{Feng2025}

\end{document}